\documentclass[12pt]{elsart} 
\usepackage{afterpage}
\usepackage{lscape}
\usepackage{amsmath}
\usepackage{amssymb}
\usepackage{setspace} 
\usepackage[dvips]{graphicx,color}

\begin{document}
\DeclareGraphicsExtensions{.eps, .jpg}
\setlength{\voffset}{-0.25in}

\begin{frontmatter} 

\title{Simple Front End Electronics for \\ Multigap Resistive Plate Chambers} 
\author[Rice]{W.J. Llope,}$^{,*}$~
\author[Rice]{T. Nussbaum,}
\author[Rice]{G. Eppley,}
\author[VU]{J. Velkovska,}
\author[VU]{T. Chujo,}
\author[VU]{S. Huang,}
\author[VU]{B. Love,}
\author[VU]{H. Valle,}
\author[BNL]{L. Ruan,}
\author[BNL]{Z. Xu,}
\author[Rice]{and B. Bonner.}

\address[Rice]{T.W. Bonner Nuclear Laboratory, Rice University, Houston, TX 77005} 
\address[VU]{Vanderbilt University, Nashville, TN 37235}
\address[BNL]{Brookhaven National Laboratory, Upton, NY 11973}

\begin{abstract} 
A simple circuit for the presentation of the signals from Multi-gap
Resistive Plate Chambers (MRPCs) to standard existing digitization electronics
is described. The circuit is based on
``off-the-shelf" discrete components. An optimization of the values of 
specific components is required to match the aspects of the MRPCs for
the given application. This simple circuit is an attractive option for
the initial signal processing for MRPC prototyping and bench- or beam-testing 
efforts, as well as for final implementations of small-area Time-of-Flight 
systems with existing data acquisition systems.
\end{abstract}

\end{frontmatter} 


$^*$ Corresponding author. Tel.: +1-713-348-4741, Fax.: +1-713-348-5215. \\
{\it E-mail address:} llope@physics.rice.edu (W.J. Llope). 

\pagebreak
\section{Introduction\label{sec:intro}}

The Time-of-Flight (TOF) systems for the ALICE \cite{ref:alicetof} and STAR
\cite{ref:startof} experiments are large-area systems based on
Multigap Resistive Plate Chambers (MRPCs). The large channel counts
of these systems demand that the front-end electronics and
digitization are done ``on-detector." The electronics for these systems
are based on the NINO chip \cite{ref:nino} for the amplification and the 
HPTDC chip \cite{ref:hptdc} for the digitization. For smaller TOF systems, 
as well as MRPC prototype testing with cosmic rays or at test-beams, simpler
front-end electronics read-out by an existing data acquisition system
can be a productive approach. Such 
electronics are inexpensive, simple to develop, low power, and 
quick and easy to construct out-of-house. An example of such electronics, 
which properly amplify the very small signals from MRPCs and drive 50 $\Omega$ signal
cables to the digitizers, is described here. This circuit
is based on readily-available ``off-the-shelf" discrete components.

Boards of this type are presently in use as the front-end electronics
for the new PHENIX ``TOF-west" system \cite{ref:tofwest}, as well as
the prototype Muon Telescope Detector (MTD) \cite{ref:mtd} in the STAR
experiment, both at the Relativistic Heavy-Ion Collider (RHIC) at
Brookhaven National Laboratory (BNL). An earlier version of this
circuit was used for the STAR TOF prototype ``TOFr" \cite{ref:tofr} in RHIC Run 3
and in a test-beam, as well as
presently in the cosmic test stands used to qualify newly constructed
STAR TOF MRPCs at the detector fabrication sites in China
\cite{ref:china}.

MRPCs are highly capacitive detectors that produce very small
signals ($\sim$10-20 fC). In order to produce output signals large
enough for digitization after long cables in typical data acquisition
systems, significant pre-amplification and subsequent amplification
of the signals is needed. The overall current gain of these electronics
into a 50 $\Omega$ load is a factor of $\sim$500. This large amount of
amplification assumes a careful and complete shielding of external
radio-frequency noise. To achieve this shielding, these electronics
exist as two layers of electronics boards. A ``feed-through" (FT)
board and a Front-End Electronics (FEE) board. The FT board bolts to
the exterior of the aluminum gas volume that encloses the MRPCs in 
a way that insures a low-impedance connection between the enclosure
ground and the electronics ground. The FT boards thus 
serve to close the MRPC gas volume as well as complete a Faraday 
cage around the MRPCs. The FT boards 
pass the detector signals to the FEE boards. 

An air gap
between the FT and FEE boards insures that the power dissipated in the
FEE board does not radiatively heat the MRPCs inside the gas volume,
which would increase the MRPC noise rates and high voltage currents.
The capacitance of MRPCs is an order of magnitude larger than that for
the typical industrial applications of the pre-amplifiers used in these electronics.
This requires an optimization of the ``matching" of these
electronics to the MRPCs. Also, 
the feed-back circuitry around
the amplifier controls the overall gain and bandwidth (signal
rise-time) and requires an optimization tailored to the specific 
application.
These optimizations are accomplished via the selection of the 
values of specific resistors in the circuit. In practice, prototype versions 
of these electronics were constructed and then tested with cosmics or test-beams
using the specific MRPCs for a given application. Different
channels of these prototype boards were configured to use different values 
for these resistors. This allowed the collection of the test data needed to
optimize these settings. Once the preferred settings were
determined, the final board production was begun.

\begin{figure}[htb]
\begin{center}
\begin{minipage}[t]{0.98\textwidth}
\begin{center}
\includegraphics[width=0.98\textwidth,keepaspectratio]{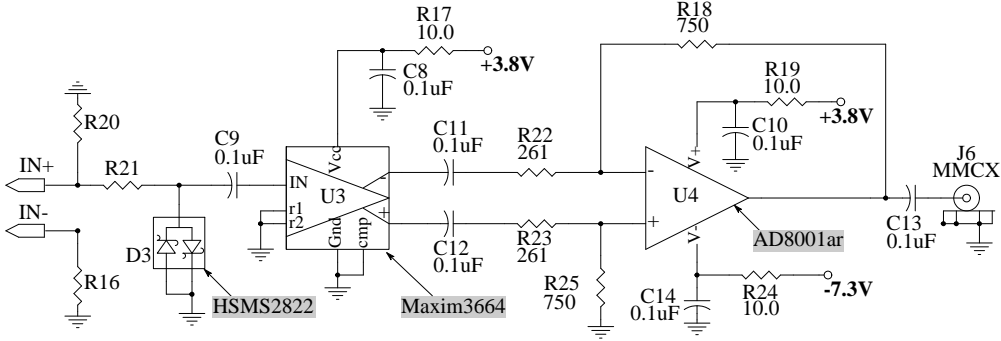}
\vspace*{-4mm}
\caption{\it The components of the present circuit.}\label{fig:circuit}
\end{center}
\end{minipage}
\end{center}
\end{figure}

The basic circuit of the FEE board is shown in Figure
\ref{fig:circuit}. Each FT and FEE board handles a total of eight MRPC
read-out channels. The MRPC signals are brought from the MRPC pads to
the underside (inside the gas volume) of the FT boards via
twisted-pair ribbon cables. ``Samtec" connectors \cite{ref:samtec}
were used to bring the MRPC signals to the inputs of the FEE boards. These 
``low-profile" connectors were
required due to the very tight integration volume for the PHENIX TOF-west
implementation. The vertical spacing between the top of the FT board
and the underside of the FEE board is 3/16" and is controlled by
threaded hex stand-offs on the 4-40 ``PEM stud" threaded posts mounted on the detector
gas volume. These same low-profile connectors were also used for the STAR MTD
prototype. 

The MRPC signals enter the FEE boards on the left side of
Figure \ref{fig:circuit}. The input labelled ``IN$+$"(``IN$-$") is connected
to the side of the MRPC to which positive(negative) high voltage
is applied. This results in negative MRPC signals at the inputs of
the FEE boards. The FEE output signals, which are also negative, 
are available at the right side of this figure. Each FEE board has six 
planes - two ground planes, two voltage planes, and two routing planes, 
although only one routing plane was actually used. 

The first components, the resistors R16, R20, and R21, deal with the
extremely reactive behavior of the MRPCs and their twisted pair signal
leads as a function of the (Fourier) frequency of the input pulses. At
``low" frequencies, the input is capacitive, while at ``high"
frequencies, the input is inductive. By design, the pre-amplifier used
here prefers input capacitances below 1 pF, which is at least an order
of magnitude below that for typical MRPCs.  
While these resistors do attenuate the
input signals, Êthe desired effect is to make the MRPC and its twisted
pair signal cables appear less reactive at the input to the pre-amplifier
to reduce its instability and ringing.  The residual
ringing still due to the internal feedback design of the pre-amplifier
was judged acceptable in these applications as the pulse area (via Analog
to Digital Converters in the digitization system), Ênot its ``time
over threshold," is used as the independent variable for the offline
pulse slewing \cite{ref:slewing} corrections. 

The resistors R16, R20, and R21 also properly terminate the
twisted-pair signal cables coming from the MRPCs. As MRPCs are
capacitive, the smallest possible impedance of the signal cabling and
the front-end electronics is desirable as it results in the fastest
rise-times of the output pulses. Thus, two twisted-pairs per MRPC 
read-out channel were used to bring the signals to the FT boards. 
Additional details on the optimization of these resistor values 
are provided below.

The next component, labelled D3, is the Schottky diode ``hsms2822" from 
Agilent, which provides the input voltage protection to the rest of the circuit. 
Input pulse heights above $\sim$400 mV are cropped at this height. 
This effectively never occurs, as this value is two orders
of magnitude higher than the detector pulse heights.

The detector signal then enters a Maxim 3664 transimpedance amplifier (the 
``pre-amplifier") which converts the current pulse from the detector to a 
differential voltage proportional to the input pulse current. 
The advantage of the Maxim 3664 transimpedance pre-amplifier is its 
much higher and more consistent ($\pm$10\%) gain compared to that for
Monolithic Microwave Integrated Circuit (MMIC) pre-amplifiers, which were
also considered for the present circuit.
The Maxim 3664 reduces the number of stages and the channel-to-channel gain variations
via the feedback design. 

\begin{figure}[htb]
\begin{center}
\begin{minipage}[t]{0.8\textwidth}
\begin{center}
\includegraphics[width=0.8\textwidth,keepaspectratio]{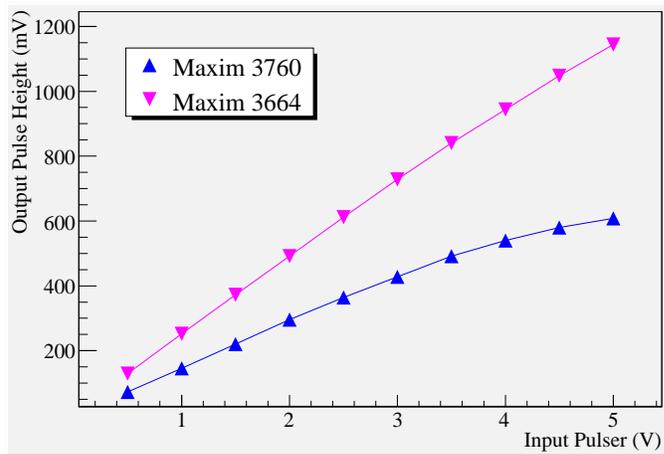}
\vspace*{-4mm}
\caption{\it The output versus (attenuated) input pulse height for the pre-amplifiers used
in the prototype and final electronics.}\label{fig:feeplot}
\end{center}
\end{minipage}
\end{center}
\end{figure}

A comparison of the gain and linearity of the FEE boards with two
different transimpedance pre-amplifier chips is shown in Figure \ref{fig:feeplot}. 
The horizontal axis is the pulse height from a pulser. The pulser
output is heavily attenuated ($\sim$40 dB) before it is input to
the FEE boards. The upward(downward) arrows indicate the pulse
height output from the FEE board when using a Maxim 3760(Maxim 3664)
pre-amplifier. The Maxim 3664 results in a higher gain and an 
improved linearity versus the input pulse height compared to the
Maxim 3760.

The differential voltage from the Maxim 3664 pre-amplifier 
is then sent to an Analog Devices AD8001ar operational amplifier. This
drives the output signals to be sent over long coaxial cables to the digitizers. 
The resistors R18 and R22 \& R23 are part of the ``feedback" network for this amplifier.
The ratio of the values of R18 to R22 \& R23 changes the overall output gain, while the 
absolute value of the resistor R18 changes the bandwidth (rise-time) of the output. 
The shortest achievable output rise-times are clearly of primary interest for the ultimate
timing of the detectors, especially in the presence of long cables before
the digitizers.

The pre-amplifier requires only a positive input voltage, while the
amplifier requires both positive and negative voltages. The rise time
of the pulses out of the amplifier are the fastest when the absolute
difference between the positive and negative voltages is on the order
of 10 V. The best results were obtained by setting the pre-amplifier
voltage to $+$3.8V, and the amplifier voltages to $+$3.8V and $-$7.3V.
These bipolar asymmetrical power supplies also minimized the power
dissipation in the voltage regulators (not shown in Figure
\ref{fig:circuit}). The currents drawn per board are approximately 240
mA and 40 mA on the positive and negative inputs, respectively. In
accordance with typical safety requirements at BNL, all power inputs
are independently protected with 0.75A and 0.3A fuses, respectively.

The resistor (10$\Omega$) and capacitor (0.1 $\mu$F) pairs R17 \& C8, 
R19 \& C10, and R24 \& C14, provide the power supply filtering for
the suppression of cross-talk across read-out channels. 
The timing cross-talk for these boards was measured
with a pulser and an 8 Gs/s Hewlett-Packard Infinium oscilloscope and was seen to 
be less than 10 ps ({\it i.e.} lower than experimentally measurable). 

Finally, the capacitor C13 ``AC-couples" the front-end electronics to the digitization
electronics. This eliminates the possible degradations to the 
performance from ground loops in the final implementation of these
electronics. The output is the ``MMCX" connector, J6. This connector was chosen 
because of its very small size (and PHENIX's very tight integration volume). 
An RG-316 type coaxial cable connects to this location
and brings the outputs to the digitization electronics for the pulse
area and timing measurements.

\begin{figure}[htb]
\begin{center}
\begin{minipage}[t]{0.8\textwidth}
\begin{center}
\includegraphics[width=0.8\textwidth,keepaspectratio]{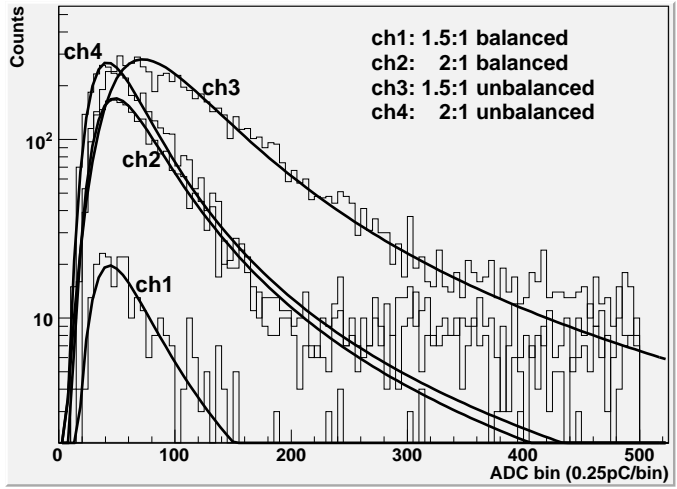}
\vspace*{-4mm}
\caption{\it The pulse-area distributions for a number of values
of the impedance-matching resistors R16, R20, and R21 obtained 
with a PHENIX TOF-west prototype MRPC.}\label{fig:adc}
\end{center}
\end{minipage}
\end{center}
\end{figure}

Once the basic circuit was developed, a few prototype boards were
built and delivered. In different channels of these prototype boards, 
different values of
the initial ``impedance matching resistors" R16, R20, and R21 were used.
Data was then collected with the PHENIX TOF-west MRPCs and the 
resulting gain of the FEE compared across these test channels. The result 
is shown in Figure \ref{fig:adc}. In the ``balanced" configurations, 
the value of the resistor R16 was set via $1/R_{16}$$=$1/$R_{20}$$+$1/$R_{21}$,
while in the ``unbalanced" configurations, the resistor R16 is removed
and this side of the MRPC output (IN-) goes directly to ground. The labels
``1.5:1" and ``2:1" denote the attenuation of the MRPC signal current from
the IN+ side of the MRPC, which is given by the ratio [$R_{20}$+$R_{21}$]/$R_{20}$.
On the basis of Figure \ref{fig:adc}, the ``1.5:1 unbalanced" configuration
was chosen, as this resulted in the largest mean-value of the
digitized pulse areas across the various test channels. The 
resistor values thus chosen for
the final production were $R_{16}$ $=$ 0, $R_{20}$ $=$ 174 $\Omega$, and
$R_{21}$ $=$ 75 $\Omega$.

Using an attenuated pulser with a $\sim$0.3 ns rise-time, the 
10\%-90\% rise-time of the output signals is 1.35 ns. 
The ``peaking time" is 2 ns.

In summary, a simple circuit based on readily-available discrete
components was developed for MRPC detectors. These boards are simple
enough that they can easily and reliably assembled ``out-of-house".
These electronics are presently in use in the PHENIX TOF-west system
and the STAR Muon Telescope Detector prototype. The ultimate timing
performance of these systems following the offline corrections, which
will be described elsewhere, met expectations.


\ack 
We gratefully acknowledge funding from the US Department of Energy under Grant
numbers DE-FG03-96ER40772 (Rice), DE-FG02-04ER41333 (Vanderbilt),
and BNL LDRD-07-007 (MTD).

 
\end{document}